\documentclass[aps,prl,twocolumn,superscriptaddress,showpacs,preprintnumbers,amsmath,amssymb,showkeys]{revtex4-1}
\usepackage{graphicx}
\usepackage{dcolumn}
\usepackage{bm}
\usepackage{notes2bib}
\usepackage{graphicx}
\usepackage{soul}
\usepackage{gensymb}
\usepackage{url}
\newcommand{\bfq}{\mbox{\boldmath $q$}}
\newcommand{\bfP}{\mbox{\boldmath $P$}}

\newcommand{\bfR}{\mbox{\boldmath $R$}}

\usepackage{amsmath}
\usepackage{latexsym}
\usepackage{amssymb}
\usepackage{color}
\usepackage[normalem]{ulem}
\usepackage{slashed}

\newcommand*{\ANL}{Argonne National Laboratory, Argonne, Illinois 60439}

\newcommand*{\SACLAY}{IRFU, CEA, Universit\'{e} Paris-Saclay, F-91191 Gif-sur-Yvette, France}

\newcommand*{\UCONN}{University of Connecticut, Storrs, Connecticut 06269}

\newcommand*{\DUKE}{Duke University, Durham, North Carolina 27708-0305}

\newcommand*{\DUQUESNE}{Duquesne University, 600 Forbes Avenue, Pittsburgh, PA 15282 }

\newcommand*{\FU}{Fairfield University, Fairfield CT 06824}

\newcommand*{\FERRARAU}{Universit\`{a} di Ferrara , 44121 Ferrara, Italy}

\newcommand*{\FIU}{Florida International University, Miami, Florida 33199}

\newcommand*{\FSU}{Florida State University, Tallahassee, Florida 32306}

\newcommand*{\GWUI}{The George Washington University, Washington, DC 20052}

\newcommand*{\ISU}{Idaho State University, Pocatello, Idaho 83209}

\newcommand*{\INFNFE}{INFN, Sezione di Ferrara, 44100 Ferrara, Italy}

\newcommand*{\INFNFR}{INFN, Laboratori Nazionali di Frascati, 00044 Frascati, Italy}

\newcommand*{\INFNGE}{INFN, Sezione di Genova, 16146 Genova, Italy}

\newcommand*{\INFNRO}{INFN, Sezione di Roma Tor Vergata, 00133 Rome, Italy}

\newcommand*{\INFNTUR}{INFN, Sezione di Torino, 10125 Torino, Italy}

\newcommand*{\INFNPAV}{INFN, Sezione di Pavia, 27100 Pavia, Italy}

\newcommand*{\ORSAY}{Universit\'{e} Paris-Saclay, CNRS/IN2P3, IJCLab, 91405 Orsay, France}

\newcommand*{\Juelich}{Institute fur Kernphysik (Juelich), Juelich, Germany}

\newcommand*{\JMU}{James Madison University, Harrisonburg, Virginia 22807}

\newcommand*{\KNU}{Kyungpook National University, Daegu 41566, Republic of Korea}

\newcommand*{\LAMAR}{Lamar University, 4400 MLK Blvd, PO Box 10046, Beaumont, Texas 77710}

\newcommand*{\MIT}{Massachusetts Institute of Technology, Cambridge, Massachusetts  02139-4307}

\newcommand*{\MISS}{Mississippi State University, Mississippi State, MS 39762-5167}

\newcommand*{\ITEP}{National Research Centre Kurchatov Institute - ITEP, Moscow, 117259, Russia}

\newcommand*{\UNH}{University of New Hampshire, Durham, New Hampshire 03824-3568}

\newcommand*{\NMSU}{New Mexico State University, PO Box 30001, Las Cruces, NM 88003, USA}

\newcommand*{\NSU}{Norfolk State University, Norfolk, Virginia 23504}

\newcommand*{\OHIOU}{Ohio University, Athens, Ohio  45701}

\newcommand*{\ODU}{Old Dominion University, Norfolk, Virginia 23529}

\newcommand*{\JLUGiessen}{II. Physikalisches Institut der Universit\"at
Gie{\ss}en, 35392 Gie{\ss}en, Germany}

\newcommand*{\URICH}{University of Richmond, Richmond, Virginia 23173}

\newcommand*{\ROMAII}{Universit\`{a} di Roma Tor Vergata, 00133 Rome Italy}

\newcommand*{\MSU}{Skobeltsyn Institute of Nuclear Physics, Lomonosov Moscow State University, 119234 Moscow, Russia}

\newcommand*{\SCAROLINA}{University of South Carolina, Columbia, South Carolina 29208}

\newcommand*{\TEMPLE}{Temple University,  Philadelphia, PA 19122 }

\newcommand*{\JLAB}{Thomas Jefferson National Accelerator Facility, Newport News, Virginia 23606}

\newcommand*{\UTFSM}{Universidad T\'{e}cnica Federico Santa Mar\'{i}a, Casilla 110-V Valpara\'{i}so, Chile}

\newcommand*{\INSUBRIA}{Universit\`{a} degli Studi dell'Insubria, 22100 Como, Italy}

\newcommand*{\BRESCIA}{Universit\`{a} degli Studi di Brescia, 25123 Brescia, Italy}

\newcommand*{\GLASGOW}{University of Glasgow, Glasgow G12 8QQ, United Kingdom}

\newcommand*{\YORK}{University of York, York YO10 5DD, United Kingdom}

\newcommand*{\VIRGINIA}{University of Virginia, Charlottesville, Virginia 22901}

\newcommand*{\WM}{College of William \& Mary, Williamsburg, Virginia 23187-8795}

\newcommand*{\YEREVAN}{Yerevan Physics Institute, 375036 Yerevan, Armenia}

\newcommand*{\CNU}{Christopher Newport University, Newport News, Virginia 23606}

\newcommand*{\NOWBRESCIA}{Universit\`{a} degli Studi di Brescia, 25123 Brescia, Italy}
{}

\begin{document}
\title{Observation of Beam Spin Asymmetries in the Process $e p\rightarrow e’\pi^+\pi^-X$ with CLAS12}

\author{T.B.~Hayward}
\affiliation{\WM}
\author{C.~Dilks}
\affiliation{\DUKE}
\author{A.~Vossen}
\affiliation{\DUKE}
\affiliation{\JLAB}
\author{H.~Avakian}
\affiliation{\JLAB}
\author {S.~Adhikari} 
\affiliation{\FIU}
\author {G.~Angelini} 
\affiliation{\GWUI}
\author{M.~Arratia}
\affiliation{University of California, Riverside, Riverside, California, 92521}
\affiliation{\JLAB}
\author {H.~Atac} 
\affiliation{\TEMPLE}
\author {C.~Ayerbe~Gayoso} 
\affiliation{\WM}
\author {N.A.~Baltzell} 
\affiliation{\JLAB}
% ...

\collaboration{The CLAS Collaboration}

\date{\today}
\begin{abstract}
The observation of beam spin asymmetries in two-pion production in semi-inclusive deep inelastic scattering off an unpolarized proton target is reported. The data presented here were taken in the fall of 2018 with the CLAS12 spectrometer using a 10.6 GeV longitudinally spin-polarized electron beam delivered by CEBAF at JLab.
The measured asymmetries provide the first opportunity to extract the parton distribution function $e(x)$, which provides information about the interaction between gluons and quarks, in a collinear framework that offers cleaner access than previous measurements. 
The asymmetries also constitute the first ever signal sensitive to the helicity-dependent two-pion fragmentation function $G_1^\perp$. A clear sign change is observed around the $\rho$ mass that appears in model calculations and is indicative of the dependence of the produced pions on the helicity of the fragmenting quark.
\end{abstract} 
\setcounter{footnote}{0}
\maketitle

Protons and neutrons constitute most of the visible matter of the universe, however our understanding of how some of their most important properties, such as mass and spin, emerge from the strong interactions of the constituent quarks and gluons is still incomplete. 
Therefore, the study of the internal dynamics of the nucleon is fundamental to our understanding of the theory of strong interactions and, by extension, our understanding of the nature of matter itself.

Parton distribution functions (PDFs) encode information about the momentum-dependent distribution of quarks inside the proton.
A PDF that is not suppressed in the cross-section by the hard scale of the process is said to be at leading twist, or twist-2~\cite{Jaffe:1997vlv} and can be interpreted as a probability distribution of the respective parton type. PDFs can also be defined for the cases including additional gluon emission or absorption by the parton after scattering. Such PDFs are in general kinematically suppressed and said to be at subleading, or higher twist.

Comparably, the non-perturbative dynamics of hadronization, the process of the formation of hadrons out of quarks and gluons, are described by fragmentation functions (FFs), which at leading twist can be interpreted in the parton model as the probability that a quark forms a certain hadron. For recent reviews, see Refs.~\cite{Aidala:2012mv,Metz:2016swz,Anselmino:2020vlp, Avakian:2019drf}.

In order to access PDFs and FFs, we consider the semi-inclusive deep inelastic scattering (SIDIS) process, where an electron scatters off a proton target at a high enough energy such that it can be described by the scattering off a single parton in the target~\cite{Anselmino:2020vlp}. This Letter reports the measurement of beam spin asymmetries for the two-pion production process in SIDIS,
\begin{equation}
\label{eq:process}
e(\ell) + p(P) \to e'(\ell') + \pi^+(P_1)+\pi^-(P_2) + X,
\end{equation}
where the quantities in the parentheses denote the respective four-momenta; boldface symbols will indicate the corresponding three-momenta. 
Fragmentation into two pions offers  more targeted access to the nucleon structure and allows for the observation of more complex phenomena in fragmentation than single-pion production~\cite{Metz:2016swz}. 

Insights into the interaction between gluons and the struck quark in the nucleon can be gained from subleading-twist PDFs. One such quantity is the collinear twist-3 PDF $e(x)$~\cite{Efremov:2002qh,Seng:2018wwp}. 
While $e(x)$ itself lacks a straightforward probabilistic interpretation, its moments provide insight: 
the first $x$-moment of $e(x)$ is related to the pion-nucleon $\sigma$-term, representing the contribution to the nucleon mass from the finite quark masses \cite{Efremov:2002qh,Courtoy:2014ixa,Ji:1994av}, and the third $x$-moment is proportional to the transverse force experienced by a transversely polarized quark in an unpolarized nucleon immediately after scattering ~\cite{Burkardt:2008ps, Pasquini:2018oyz}.
Like the other collinear PDFs, $e(x)$ is dependent on the scaling variable $x$, which in the parton picture corresponds to the light-cone momentum fraction carried by the probed quark~\cite{Jaffe:1990qh,Anselmino:2020vlp} and can be expressed as $x=Q^2/(2P^\mu q_\mu)$. As usual $Q^2=-q^\mu q_\mu$ denotes the scale of the process, where $q=\ell-\ell'$ is the four-momentum of the exchanged virtual photon.

A first model-dependent extraction of $e(x)$ from single-hadron data has been performed~\cite{Efremov:2002ut}, along with another extraction from preliminary two-pion data from CLAS~\cite{Mirazita:2020lik, Courtoy:2014ixa}. In SIDIS single-hadron production, $e(x)$ can only be accessed via beam spin asymmetries with the inclusion of the transverse momentum dependence (TMD) of the FF. This leads to a convolution of the PDF and FF over the TMD. Furthermore, factorization of the cross section into PDFs and FFs in the TMD framework is not yet proven at subleading twist~\cite{Bacchetta:2019qkv}. These issues motivate the high-precision measurement of two-pion beam spin asymmetries presented here.

In addition to $e(x)$, the other primary focus of the presented measurements is on the dihadron FF $G_1^\perp$, which describes the dependence of two-pion production on the helicity of the fragmenting quark. No previous measurement sensitive to $G_1^\perp$ exists.
Recently, interest in the possible mechanism behind $G_1^\perp$ led to several model calculations~\cite{Matevosyan:2017alv,Luo:2020wsg}. 
In Ref.~\cite{Luo:2020wsg} interference between different partial waves leads to a signal with a distinct dependence on the two-pion invariant mass $M_h$, with a sign change around the $\rho$ mass. It is also interesting to note that $G_1^\perp$ could be sensitive to QCD vacuum fluctuations~\cite{Boer:2003ya} and thus to the strong CP problem. 

FFs describing two-pion production depend on $M_h$ and on $z$, the fraction of the fragmenting quark momentum carried by the pion pair. Dihadron FFs can be decomposed into partial waves~\cite{Bacchetta:2002ux,Gliske:2014wba}, with the corresponding associated Legendre polynomials depending on the angle $\theta$ between the hadron momentum $\bfP_1$ in the dihadron center-of-mass frame, and the direction of the pair momentum $\bfP_h$ in the photon-target rest frame. This dependence is integrated over the CLAS12 acceptance in the results shown here and the relevant mean $\theta$ values are given in the ancillary files.

The data were taken with the CLAS12 spectrometer~\cite{Burkert:2020akg} using a 10.6 GeV longitudinally polarized electron beam delivered by CEBAF, incident on a liquid-hydrogen target. The beam polarization averaged to $86.9\%\pm 2.6\%$ and was flipped at 30~Hz to minimize systematic effects. 
This analysis uses the Forward Detector of CLAS12, which contains a tracking subsystem consisting of drift chambers in a toroidal magnetic field and high and low-threshold Cherenkov counters to identify the scattered electron and final state pions. Additional identification is performed for electrons with an electromagnetic calorimeter and for pions by six arrays of plastic scintillation counters.

SIDIS events were selected by requiring $Q^2~>~1$~GeV$^2$ and the mass of the hadronic final state to be above $2$~GeV. Exclusive reactions were removed  with the condition on the missing mass $M_X~>~1.5$~GeV, defined as the mass of the unmeasured part of the final state.
Contributions from events where a photon is radiated from the incoming lepton were reduced by placing a condition of $y<0.8$, where $y=P^\mu q_\mu/(P^\mu l_\mu)$ is the fractional energy loss of the scattered electron, and by requiring a minimum momentum of 1.25~GeV for each pion.
Finally, contributions from the target fragmentation region were reduced by requiring $x_F>0$ for each pion,
where $x_F$ denotes Feynman-$x$ and takes a positive value if the outgoing hadron moves in the same direction as the incoming electron, in the struck quark center-of-mass frame.

\begin{figure}
    \centering
    \includegraphics[width=0.45\textwidth]{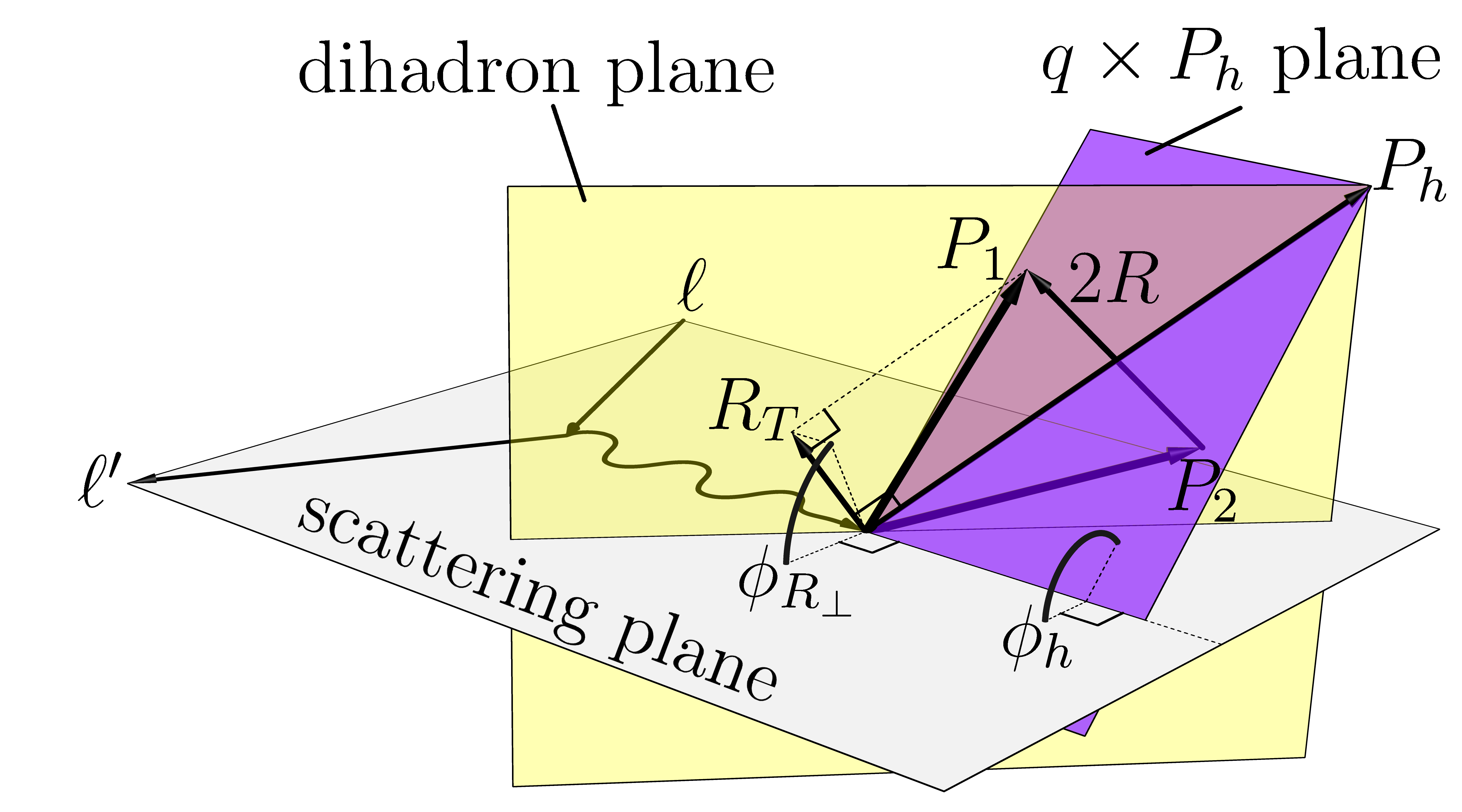}
    \caption{The coordinate system used in this analysis. The electron scattering plane is spanned by the incoming and outgoing lepton, the dihadron plane is spanned by $\bfP_1$ and $\bfP_2$, containing also $\bfP_h$, $\bfR$, and $\bfR_T$, and the $\bfq\times\bfP_h$ plane contains only $\bfq$ and $\bfP_h$. The azimuthal angles $\phi_{h}$ and $\phi_{R_\perp}$ are defined within the plane transverse to $\bfq$, from the electron scattering plane to, respectively, the $\bfq\times\bfP_h$ plane and the dihadron plane. 
    }
    \label{fig:cooSystem}
\end{figure}

The correlations between quark and gluon fields in the nucleon encoded in $e(x)$, as well as the hadronization process described by $G_1^\perp$, are imprinted in the azimuthal angles of the final state hadrons~\cite{Bacchetta:2002ux,Bacchetta:2003vn,Gliske:2014wba}. An observable sensitive to these functions can be constructed by analyzing beam helicity-dependent azimuthal modulations of the two-pion cross section.
Figure~\ref{fig:cooSystem} illustrates the two-pion three-momenta $\bfP_h=\bfP_{1}+\bfP_{2}$ and $2 \bfR=\bfP_{1}-\bfP_{2}$, where $\bfP_1$ is assigned to the $\pi^+$. The azimuthal angles $\phi_h$ and $\phi_{R_\perp}$ are defined as \begin{align} 
\phi_{h}&= 
\frac{ \left(\bm q\times\bm l\right)\cdot \bm P_h }{| \left(\bm q\times\bm l\right)\cdot \bm P_h |}  
\arccos \frac{ \left(\bm q \times \bm l\right) \cdot \left(\bm q \times \bm P_h \right)}
                                          {\left| \bm q \times \bm l\right| \left| \bm q \times \bm P_h \right| } \; ,
\\
\phi_{R_\perp}&= 
\frac{ \left(\bm q\times\bm l\right)\cdot \bm R_T }{| \left(\bm q\times\bm l\right)\cdot \bm R_T |}  
\arccos \frac{ \left(\bm q \times \bm l\right) \cdot \left(\bm q \times \bm R_T \right)}
                                          {\left| \bm q \times \bm l\right| \left| \bm q \times \bm R_T \right| } \; ,
\label{eq:phiDef}
\end{align}
where $\bfR_T$ is the component of $\bfR$ perpendicular to $\bfP_h$, calculated as $\bfR_T=\left(z_2\bfP_1^\perp-z_1\bfP_2^\perp\right)/z$~\cite{Matevosyan:2017alv}.

The beam helicity-dependent part of the two-pion cross section can be written in terms of PDFs and FFs, integrating over partonic transverse momenta at subleading twist as~\cite{Bacchetta:2003vn,Bacchetta:2002ux,Gliske:2014wba}
\begin{align}
\label{eq:sinphir_xsec}
   & d\sigma_{LU}\propto \\ 
    &W\lambda_e \sin(\phi_{R_\perp})\left( x e(x) H_1^\sphericalangle(z,M_h)+ \frac{1}{z}f_1(x) \tilde{G}^\sphericalangle(z,M_h)  \right)\nonumber
    \\& +\ldots .\nonumber
\end{align}
Here, the subscript ${LU}$ refers to a longitudinally polarized beam and an unpolarized target, $\lambda_e$ is the electron helicity, and $W$ is a proportionality factor appropriate for twist-3 modulations and dependent on $x$ and $y$, interpreted as the depolarization of the exchanged virtual photon~\cite{Bacchetta:2002ux,Bacchetta:2003vn,Gliske:2014wba}. 
Additional azimuthal modulations exist that can also be extracted in a simultaneous fit.
Eq.~\eqref{eq:sinphir_xsec} omits the sum over quark flavors.

The dihadron FF $H_1^\sphericalangle$ that $e(x)$ is multiplied by is sensitive to the transverse polarization of the outgoing quark and has been extracted from a combined analysis of $e^+e^-$ data and Monte Carlo tuned to Belle kinematics~\cite{Vossen:2011fk,Courtoy:2012ry}. The second term contains the well-constrained unpolarized PDF $f_1(x)$ and the twist-3 dihadron FF $\tilde{G}^\sphericalangle$, which is significantly smaller than $H_1^\sphericalangle$ in model calculations~\cite{Yang:2019aan}, but remains unmeasured. 
The comparison between future target spin asymmetry measurements and the reported beam spin asymmetries may help shed light on the contributions of $\tilde{G}^\sphericalangle$~\cite{Courtoy:2014ixa}.

When the dependence on transverse momenta is included, the cross section depends on $\phi_h$ and the dihadron FF $G_1^\perp$ appears in a leading-twist term:

\begin{equation}
 \label{eq:g1perp}
d\sigma_{LU}\propto  C\lambda_e \sin(\phi_h-\phi_{R_\perp})  \mathcal{I}\left[ f_1 G_1^\perp \right] +\ldots ,
\end{equation}
where $C$ is the corresponding depolarization factor for twist-2 modulations and again  additional terms exist in the cross section.
As $G_1^\perp$ is a TMD FF, it appears in Eq.~\eqref{eq:g1perp} in a convolution, denoted by $\mathcal{I}$, of the transverse momentum dependence of $f_1(x)$, which has been constrained by data~\cite{Bacchetta:2019sam}, with that of $G_1^\perp$~\cite{Bacchetta:2002ux,Gliske:2014wba,Matevosyan:2017liq}. 

The individual terms can be extracted from Eqs.~\eqref{eq:sinphir_xsec} and \eqref{eq:g1perp} by forming the beam spin asymmetry $A_{LU}$ from the two-pion yields $N^\pm$, produced from the scattering of an electron with helicity $\pm$, written
\begin{align}
\label{eq:fit}
   & A_{LU}=
   \frac{1}{P_\text{beam}} \frac{N^+(\phi_h,\phi_{R_\perp})- N^-(\phi_h,\phi_{R_\perp})}{N^+(\phi_h,\phi_{R_\perp})+ N^-(\phi_h,\phi_{R_\perp})}=\\
     & A_{LU}^{\sin\left(\phi_h-\phi_{R_\perp}\right)}\sin(\phi_h-\phi_{R_\perp})+A_{LU}^{\sin\left(\phi_{R_\perp}\right)}\sin(\phi_{R_\perp})\nonumber\\
     & +\ldots \nonumber,
\end{align}
and fitting for the resulting azimuthal modulation amplitudes, with $P_\text{beam}$ the beam polarization. The amplitudes in Eq.~\eqref{eq:fit} were extracted from the data using an unbinned maximum likelihood fit that includes additional azimuthal modulations beyond the two listed here, from the cross section partial waves up to $\ell=2$; see Ref.~\cite{Gliske:2014wba} for details. 
A binned $\chi^2$-minimization fit with $8\times 8$ bins in $\phi_h$ and $\phi_{R_\perp}$ was also performed and is in very good agreement with the unbinned fit with a mean reduced $\chi^2$ of 1.05.
The resulting asymmetries have been divided by the polarization and can be further corrected for the ratio of the depolarization factors $W(x,y)$ and $C(x,y)$ in Eqs.~\eqref{eq:sinphir_xsec} and \eqref{eq:g1perp} to the respective factor $A(x,y)$ of the unpolarized cross section using information given in the ancillary files.

\begin{figure}
    \centering
    \includegraphics[width=0.48\textwidth]{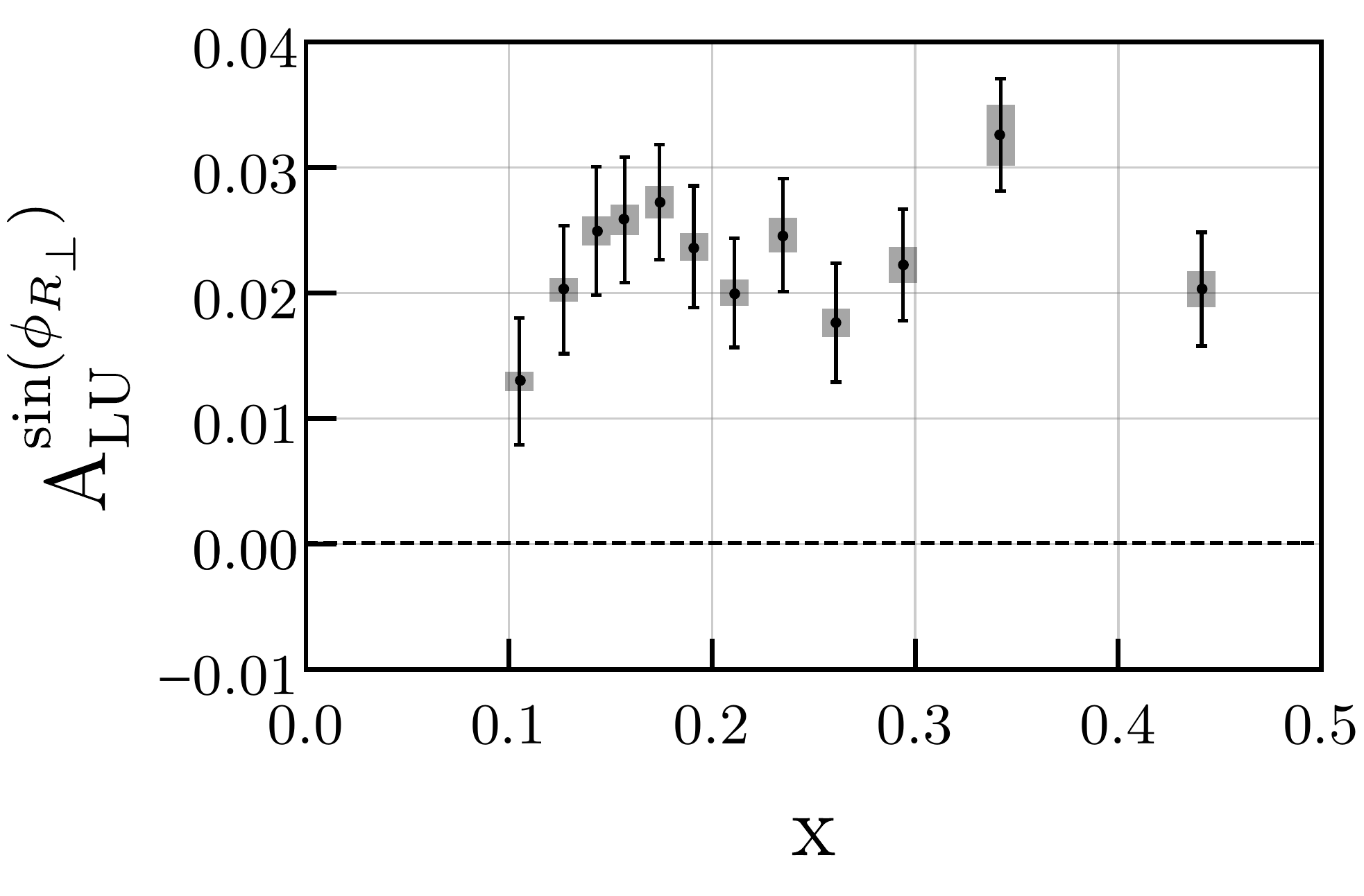}
    \caption{The measured $A_{LU}^{\sin\left(\phi_{R_\perp}\right)}$ asymmetry vs.~$x$. The thin, black bars indicate statistical uncertainties and the vertical extent of the wide, gray bars indicates systematic uncertainties.}
    \label{fig:ex}
\end{figure}

\begin{figure}
    \centering
     \includegraphics[width=0.48\textwidth]{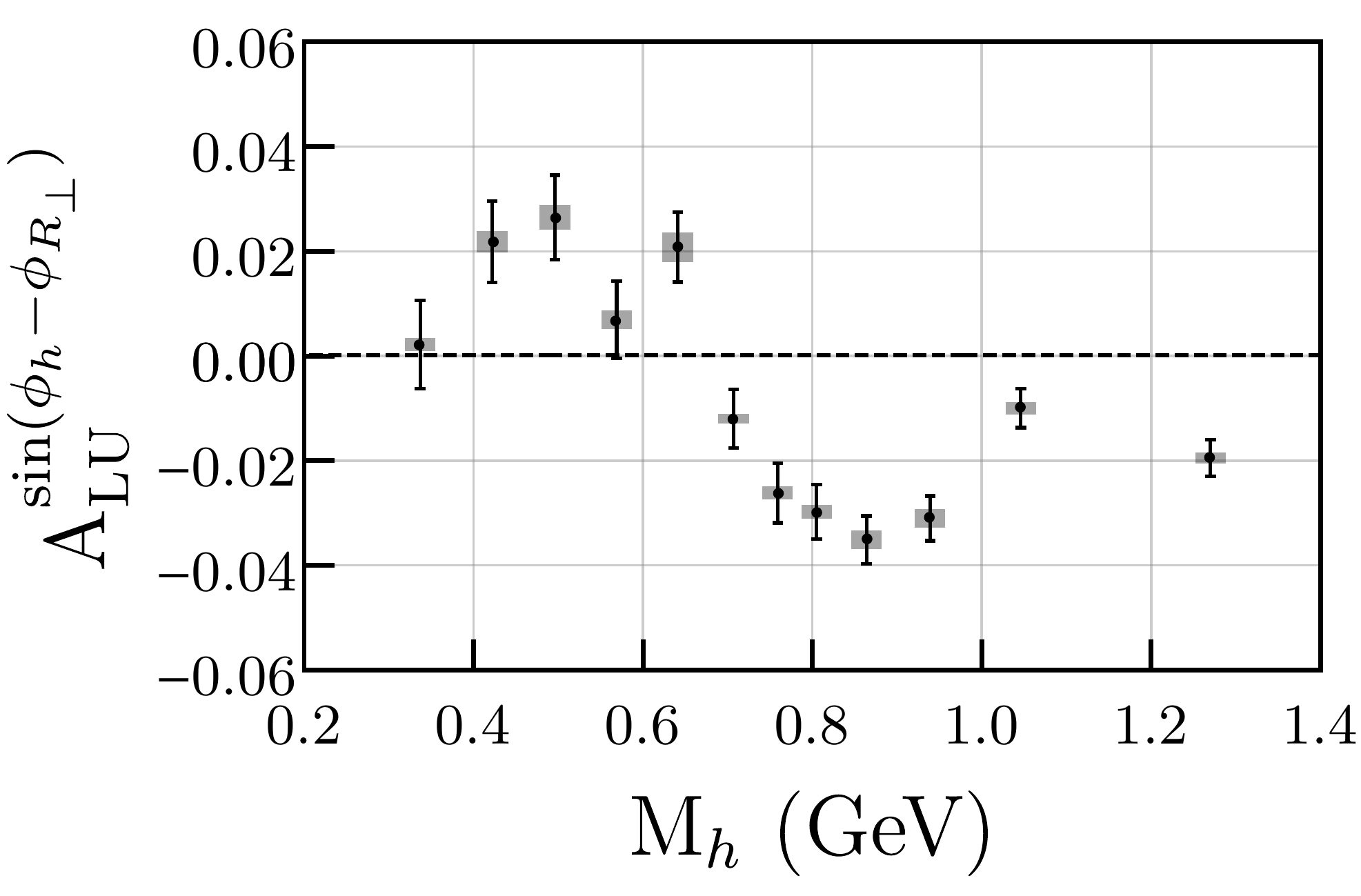}
    \caption{The measured $A_{LU}^{\sin(\phi_h - \phi_{R_\perp})}$ asymmetry vs.~$M_h$. The thin, black bars indicate statistical uncertainties and the vertical extent of the wide, gray bars indicates systematic uncertainties.}
    \label{fig:g1VsM}
\end{figure}
\begin{figure}
    \centering
    \includegraphics[width=0.49\textwidth]{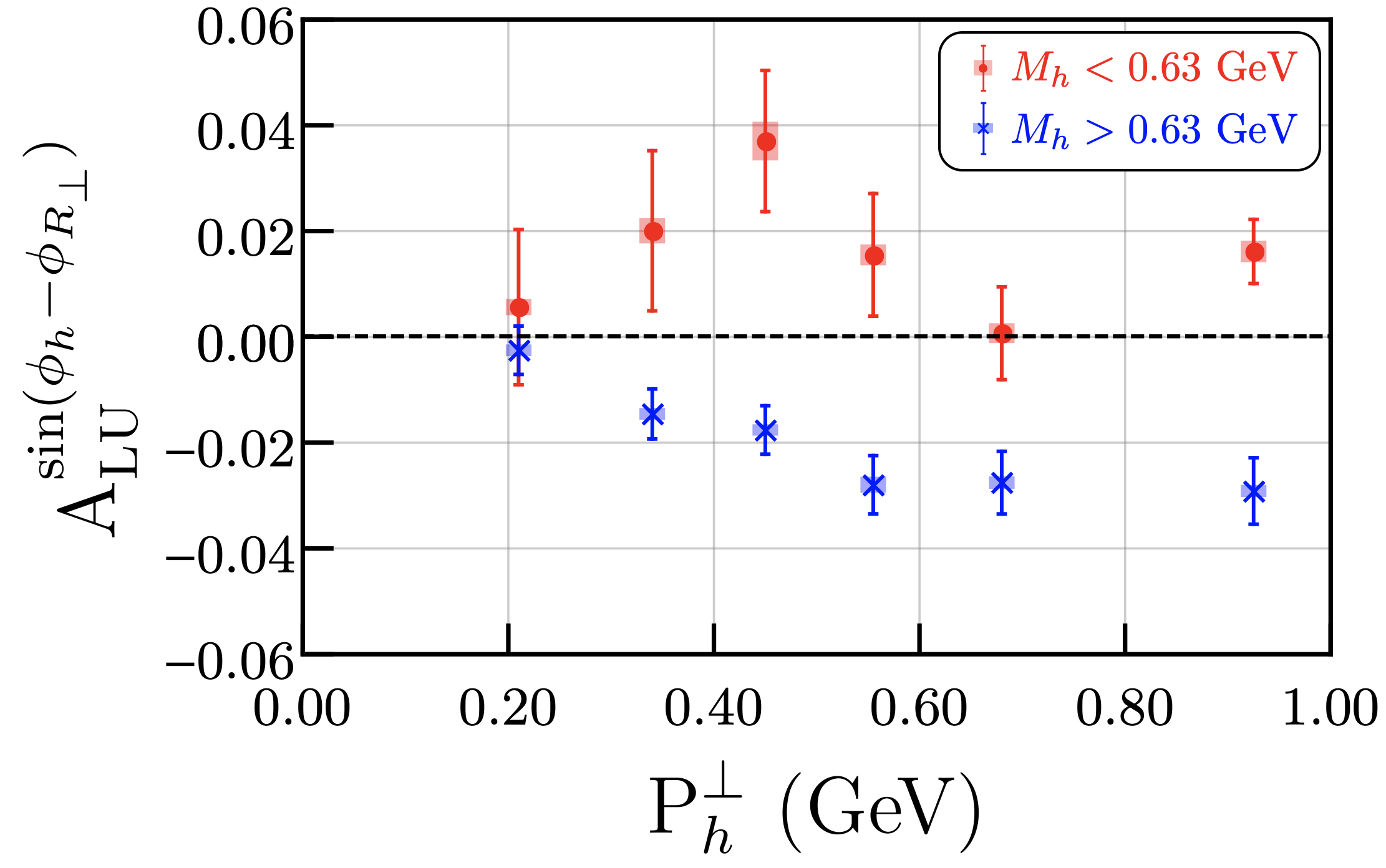}
    \caption{The measured $A_{LU}^{\sin(\phi_h - \phi_{R_\perp})}$ asymmetry vs.~$P_h^\perp$. The data have been split into two bins of $M_h$ above and below 0.63~GeV. Asymmetries for lower values of $M_h$ are shown in red circles and the blue crosses show the values for higher $M_h$. The thin, solid bars indicate statistical uncertainties and the vertical extent of the wide bars indicates systematic uncertainties.}
    \label{fig:g1VsPt}
\end{figure}
\begin{figure}
    \centering
    \includegraphics[width=0.48\textwidth]{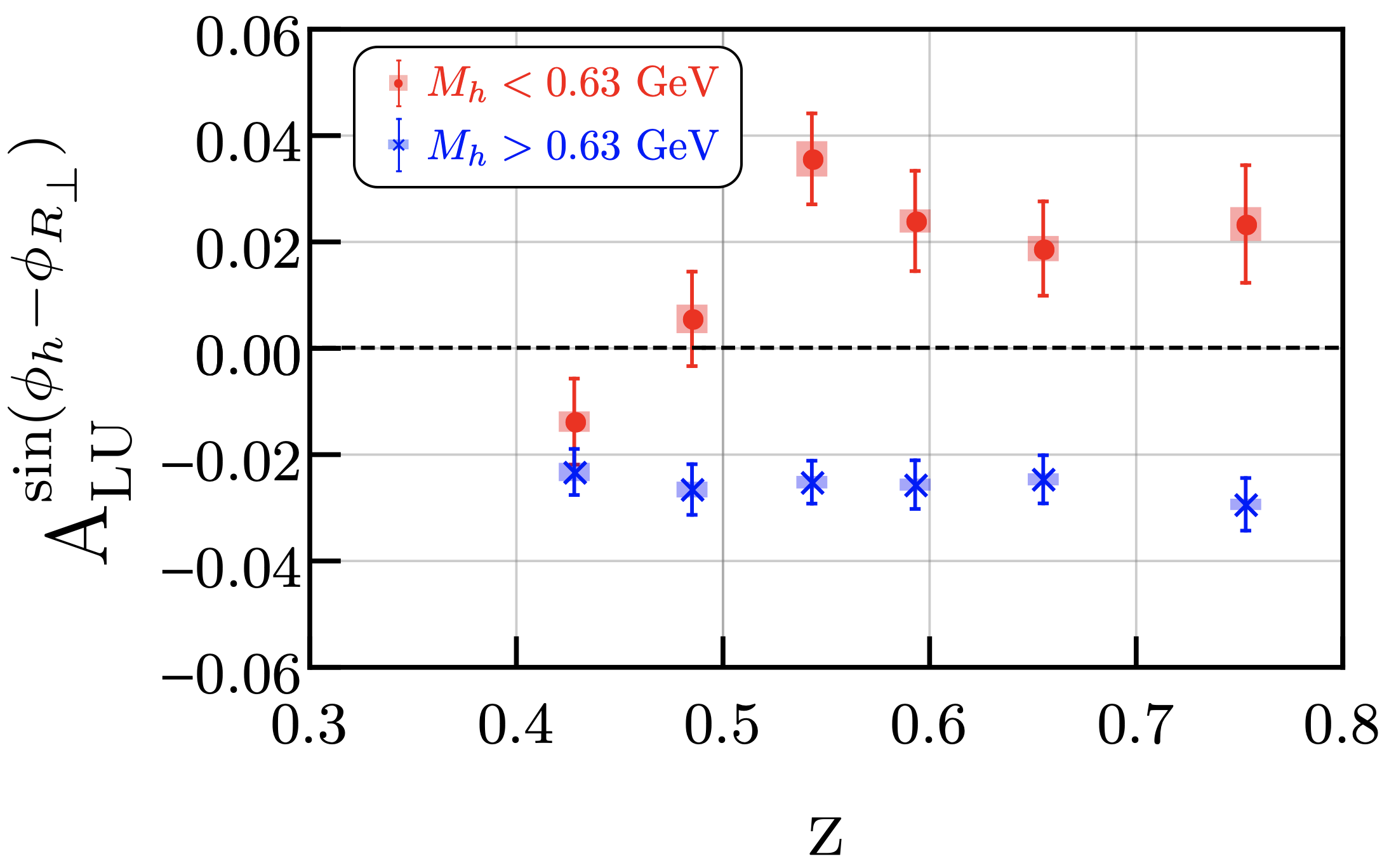}
    \caption{The measured $A_{LU}^{\sin(\phi_h - \phi_{R_\perp})}$ asymmetry vs.~$z$. The data have been split into two bins of $M_h$ above and below 0.63~GeV. Asymmetries for lower values of $M_h$ are shown in red circles and the blue crosses show the values for higher $M_h$. The thin, solid bars indicate statistical uncertainties and the vertical extent of the wide bars indicates systematic uncertainties.}
    \label{fig:g1VsZ}
\end{figure}

Figure~\ref{fig:ex} shows the result for $A_{LU}^{\sin\left(\phi_{R_\perp}\right)}$  vs.~$x$ and integrated over the other relevant variables.
A significant signal is observed that is relatively flat throughout the valence quark region. The PDF $e(x)$ is confirmed to be nonzero and its general shape can be inferred because the asymmetry presented here is proportional to $e(x)H_1^\sphericalangle(z,M_h)$ and $H_1^\sphericalangle(z,M_h)$ has been studied previously~\cite{Courtoy:2014ixa}.
The function $e(x)$ can be extracted point-by-point from these data when combined with knowledge about $H_1^\sphericalangle$ and careful consideration of the second term in Eq.~\ref{eq:sinphir_xsec}.

In Figs.~\ref{fig:g1VsM}-\ref{fig:g1VsZ} results for $A_{LU}^{\sin(\phi_h-\phi_{R_\perp})}$, sensitive to $G_1^\perp$, are shown vs.~$M_h$, $P_h^\perp$, and $z$, integrated over the other variables. 
The quantity $P_h^\perp$, the transverse momentum of the final-state pion pair with respect to $\bfq$, accesses the convolution of the TMD of the PDF and dihadron FF.
In particular, a dependence on $M_h$ with an explicit sign change around the $\rho$ mass is seen. This behavior is consistent with model calculations~\cite{Luo:2020wsg} and originates from the real part of the interference of $s$ and $p$-wave dihadrons. This significant asymmetry with its sign change is clear experimental evidence that the produced pions depend on the helicity of the fragmenting quark.

In order to investigate the possible differences in effects coming from uncorrelated and correlated hadrons, the data were further split into events with $M_h<0.63$~GeV and $M_h>0.63$~GeV to observe the dependence on $z$ and on $P_h^\perp$.
The dependence on $P_h^\perp$ is of special interest, since here for the first time results are shown that are sensitive to a TMD fragmentation into two pions. 
It is a common assumption that the transverse momentum dependence of the PDFs and FFs is Gaussian~\cite{Anselmino:2020vlp} and
the data are consistent with this assumption. One conjecture about the source of the different sign in both mass regions is that for $M_h>0.63$~GeV, vector mesons make up a significant fraction of the hadron pairs, which changes the transverse momentum spectrum. Finally, the dependence of the asymmetry on $z$, shown in Fig.~\ref{fig:g1VsZ}, is relatively flat for both $M_h$ bins with the exception of $z<0.5$ for the lower $M_h$ bin, where the asymmetry is smaller.

Systematic effects on these measurements have been studied using a Monte Carlo simulation based on the PEPSI generator~\cite{Mankiewicz:1991dp} and a GEANT4-based simulation of the detector~\cite{Agostinelli:2002hh,Ungaro:2020xlc} that was tuned to match the CLAS12 data. 
The systematic uncertainties are dominated by contributions from baryonic decays from the target fragmentation region, bin migration effects and a scale uncertainty stemming from the uncertainty on the beam polarization. 
Baryonic contributions from the target fragmentation region are dependent on $z$, reaching up to 6\% at the lowest $z$ but falling steeply to about 1\% at $z$ of 0.755. Bin migration effects are only significant for the $M_h$ dependence of $A_{LU}^{\sin(\phi_h-\phi_{R_\perp})}$, which changes rapidly around the $\rho$ mass. In this region, systematic uncertainties from bin migration reach up to 10\% of the asymmetry. The beam polarization scale uncertainty is 3.0\%.

Several additional sources of systematic uncertainties have been studied but found to be negligible. Contributions include particle identification, radiative effects, accidental coincidences and the photoproduction of electrons that are misidentified as the scattered electron. 

Eqs.~\eqref{eq:sinphir_xsec} and \eqref{eq:g1perp} show the beam spin dependent part of the cross section, however, the asymmetries $A_{LU}$ are normalized by the beam spin independent cross section $\sigma_{UU}$.  The unknown relative strength of the partial waves contributing to $\sigma_{UU}$, along with their non-orthogonality within the experimental acceptance, leads to an effective shift in the extracted asymmetries. The size of this effect has been estimated elsewhere \cite{Airapetian:2008sk}, but a precise systematic assignment requires a more thorough understanding of the unpolarized fragmentation function than is currently available.
The ancillary files contains estimates of the effect on $A_{LU}$ based on Monte Carlo studies, however these estimates are based on an assumption of the size of the yet unknown $\sigma_{UU}$ modulation amplitudes and are therefore not included in the presented systematic uncertainties. In the future, when the amplitudes of the unpolarized cross section are better constrained, it should be possible to use formulae in the ancillary files to update the asymmetry values given here to reflect the additional contributions.

In summary, this Letter reports the first significant beam spin asymmetries observed in two-pion production in SIDIS. The data indicate a non-zero signal for the azimuthal modulation sensitive to the subleading-twist PDF $e(x)$ which may enable a point-by-point extraction of this quantity. Additionally, the first measurement sensitive to $G_1^\perp$, the helicity-dependent dihadron FF, is reported. Figures~\ref{fig:ex}--\ref{fig:g1VsZ} show the main results, and all asymmetry measurements are included in the CLAS Physics Database \cite{clasDB}.
Future work will concentrate on a measurement of the partial wave decomposition of $\sigma_{LU}$ and $\sigma_{UU}$, which will address the uncertainty discussed above but is also interesting in its own right in order to gain further insight into hadronization phenomena as well.

We acknowledge the outstanding efforts of the staff of the Accelerator and the Physics Divisions at Jefferson Lab in making this experiment possible. This work was supported in part by the U.S. Department of Energy, the National Science Foundation (NSF), the Italian Istituto Nazionale di Fisica Nucleare (INFN), the French Centre National de la Recherche Scientifique (CNRS), the French Commissariat pour l$^{\prime}$Energie Atomique, the UK Science and Technology Facilities Council, the National Research Foundation (NRF) of Korea, the Helmholtz-Forschungsakademie Hessen für FAIR (HFHF) and the Ministry of Science and Higher Education of the Russian Federation. The Southeastern Universities Research Association (SURA) operates the Thomas Jefferson National Accelerator Facility for the U.S. Department of Energy under Contract No. DE-AC05-06OR23177. TH thanks the Department of Energy for support under grant DE-FG02-96ER41003. The work of CD and AV is supported by the 
U.S. Department of Energy, Office of Science, Office of Nuclear Physics under 
Award Numbers DE-SC0019230 and DE-AC05-06OR23177.

\bibliography{arxiv_submission_2}
\end{document}